\pdfoutput=1
\documentclass[10pt,UTF8]{article}
\usepackage{amssymb,amsmath,multicol,titlesec}
\usepackage[hmargin = 1in, vmargin = 1in]{geometry}

\usepackage[parfill]{parskip}
\usepackage{graphicx}
\usepackage{caption}
\usepackage{titlesec}
\usepackage{subfigure}
\usepackage{mathptmx}
\usepackage{flushend}
\usepackage{latexsym}
\usepackage{xspace}
\usepackage[colorlinks, linkcolor=black,anchorcolor=black,citecolor=black,urlcolor=black]{hyperref}
\usepackage[numbers,sort&compress]{natbib}

\titleformat{\section}[hang]{\fontsize{12pt}{1em}\selectfont \bfseries}{\thesection. }{0pt}{\centering \MakeUppercase}
\titleformat{\subsection}[hang]{\fontsize{11pt}{1em}\selectfont \bfseries}{\thesubsection}{5pt}{}
\titleformat{\subsubsection}[runin]{}{\thesubsubsection}{5pt}{}
\titlespacing{\section}{0pt}{10pt}{10pt}
\titlespacing{\subsection}{0pt}{10pt}{-5pt}
\titlespacing{\subsubsection}{0pt}{8pt}{4pt}	
\usepackage{fancyhdr}
\newcommand*{\ninoone}{\ensuremath{\mathchoice%
      {\displaystyle\raise.4ex\hbox{\(\displaystyle\tilde\chi^0_1\)}}%
         {\textstyle\raise.4ex\hbox{\(\textstyle\tilde\chi^0_1\)}}%
       {\scriptstyle\raise.3ex\hbox{\(\scriptstyle\tilde\chi^0_1\)}}%
 {\scriptscriptstyle\raise.3ex\hbox{\(\scriptscriptstyle\tilde\chi^0_1\)}}}\xspace}
\newcommand*{\chinopm}{\ensuremath{\mathchoice%
      {\displaystyle\raise.4ex\hbox{\(\displaystyle\tilde\chi^\pm\)}}%
         {\textstyle\raise.4ex\hbox{\(\textstyle\tilde\chi^\pm\)}}%
       {\scriptstyle\raise.3ex\hbox{\(\scriptstyle\tilde\chi^\pm\)}}%
 {\scriptscriptstyle\raise.3ex\hbox{\(\scriptscriptstyle\tilde\chi^\pm\)}}}\xspace}
\newcommand*{\nino}{\ensuremath{\mathchoice%
 {\displaystyle\raise.4ex\hbox{\(\displaystyle\tilde\chi^0\)}}%
    {\textstyle\raise.4ex\hbox{\(\textstyle\tilde\chi^0\)}}%
  {\scriptstyle\raise.3ex\hbox{\(\scriptstyle\tilde\chi^0\)}}%
{\scriptscriptstyle\raise.3ex\hbox{\(\scriptscriptstyle\tilde\chi^0\)}}}\xspace}
\newcommand*{\chinoonepm}{\ensuremath{\mathchoice%
      {\displaystyle\raise.4ex\hbox{\(\displaystyle\tilde\chi^\pm_1\)}}%
         {\textstyle\raise.4ex\hbox{\(\textstyle\tilde\chi^\pm_1\)}}%
       {\scriptstyle\raise.3ex\hbox{\(\scriptstyle\tilde\chi^\pm_1\)}}%
 {\scriptscriptstyle\raise.3ex\hbox{\(\scriptscriptstyle\tilde\chi^\pm_1\)}}}\xspace}
\newcommand*{\ninotwo}{\ensuremath{\mathchoice%
      {\displaystyle\raise.4ex\hbox{\(\displaystyle\tilde\chi^0_2\)}}%
         {\textstyle\raise.4ex\hbox{\(\textstyle\tilde\chi^0_2\)}}%
       {\scriptstyle\raise.3ex\hbox{\(\scriptstyle\tilde\chi^0_2\)}}%
 {\scriptscriptstyle\raise.3ex\hbox{\(\scriptscriptstyle\tilde\chi^0_2\)}}}\xspace}

\begin{document}

\begin{center}
    \vspace{0.2in}
    \noindent{\fontsize{14pt}{1em}\selectfont \textbf{Prospects for chargino pair production at CEPC}}\\[12pt]
    {\fontsize{11pt}{1.2em}\selectfont
    Jia-Rong Yuan \textsuperscript{12}, Hua-Jie Cheng \textsuperscript{13},Xu-Ai Zhuang \textsuperscript{*1}
    \\[10pt]
    \textsuperscript{1} Institute of High Energy Physics, Chinese Academy of Science, Yuquan Road 19B, Shijingshan District, Beijing 100049, China\\
    \textsuperscript{2} University of Chinese Academy of Sciences, Yuquan Road 19A, Shijingshan District, Beijing 100049, China\\
    [10pt]
    \textsuperscript{3} Department of Physics, National Taiwan University, Roosevelt Road 1 Sec. 4, Taipei 10617, Taiwan\\[10pt]
    * email:zhuangxa@ihep.ac.cn (corresponding author)\\
    [0.6in]
    }
    \end{center}
    \section*{ABSTRACT}
    The proposed Circular Electron Positron Collider (CEPC), with a center-of-mass energy $\sqrt{s} = 240$ GeV, will primarily serve as a Higgs factory.
    At the same time, it can offer good opportunities to search for new physics phenomena at low energy, which can be challenging to discover at hadron colliders, but well motivated by some theoretical models developed to explain, e.g., the relic abundance of dark matter.
    This paper presents sensitivity studies of chargino pair production, considering scenarios for both a Bino-like and a Higgsino-like neutralino as the lightest supersymmetric particle, using full Monte Carlo (MC) simulation.
    With the assumption of systematic uncertainties at the level of 5\%, the CEPC has the ability to discover chargino pair production up to the kinematic limit of $\sqrt{s}/2$ for both scenarios.
    The results have a minor dependence on the reconstruction model and detector geometry. 
    These results can also be considered as a reference and benchmark for similar searches at other proposed electron-positron colliders, such as the Future Circular Collider ee (FCC-ee) or the International Linear Collider (ILC), particularly given the similar nature of the facilities, detectors, center-of-mass energies, and target luminosities.
    \par\textbf{Keywords: }CEPC; chargino; Bino; Higgsino

\let\thefootnote\relax\footnotetext{* This study was supported by the National Key Programme (Grant NO.: 2018YFA0404000).}
\tableofcontents
\clearpage

 \section{Introduction}
 \label{sec:intro}
 Supersymmetry (SUSY)~\cite{Golfand:1971iw,Volkov:1973ix,Wess:1974tw,Wess:1974jb,Ferrara:1974pu,Salam:1974ig,Martin:1997ns} predicts new particles, each of whose spin differs by a half unit from their corresponding standard model (SM) particles. 
In SUSY models with conserved $R$-parity~\cite{Farrar:1978xj}, SUSY particles are always produced in pairs, and the lightest supersymmetric particle (LSP) is stable and can be considered as a potential dark matter candidate~\cite{Goldberg:1983nd,Ellis:1983ew}. 

The charginos $\tilde{\chi}^{\pm}_i$ $(i=1,2)$ and neutralinos $\tilde{\chi}^0_j$ $(j=1,2,3,4)$ are referred to as electroweakinos. They are the mass eigenstates formed from linear superpositions of the Bino, Wino and Higgsino particles, the superpartners of the charged and neutral Higgs bosons and electroweak gauge bosons.
The subscripts i and j indicate states of increasing mass. 
In the Minimal Supersymmetric Standard Model (MSSM), the magnitude of the Bino, Wino, and Higgsino mass parameters are referred to as $M_1$, $M_2$ and $\mu$ respectively.
The mass splitting between the electroweakinos mainly depends on the absolute values of $M_1$, $M_2$ and $\mu$.
Two SUSY scenarios are considered in this paper. 
In the first scenario, the absolute values of $M_1$ and $M_2$ parameters are considered to be near the weak scale and similar in magnitude, while the magnitude of $\mu$ is significantly larger, such that $|M_1|<|M_2|\ll |\mu|$. In this case, the LSP \ninoone is Bino-like \ninoone and the next to lightest supersymmetric particle (NLSP) is part of a Wino-like doublet forming \ninotwo and \chinoonepm.  
The second scenario considers the case that the absolute value of $\mu$ is near the weak scale, while the magnitude of $M_1$ and $M_2$ can be significantly larger, i.e. $|\mu|\ll |M_1|,|M_2|$, so that \ninoone, \chinoonepm and \ninotwo are Higgsino-like and almost mass degenerate.
The first scenario is favored by dark matter arguments and the second scenario is motivated by naturalness considerations~\cite{Kitano:2005ast,Kitano:2006sns,Baer:2013rns}, but is a challenging experimental signature to search for due to the small mass splittings between NLSP and LSP.

Direct searches for chargino pair production with Bino-like or Higgsino-like LSPs were performed previously at the Large Electron-Positron Collider (LEP) and the Large Hadron Collider (LHC).
This process was excluded by LEP for \chinoonepm masses below 92.4 (91.9) GeV independently of the \ninoone mass for the Higgsino (Bino) LSP case~\cite{LEPchargino,Heister:2002mn,Abdallah:2003xe,Chemarin:571780,CERN-OPAL-PN-464,CERN-OPAL-PN-470}.
At ATLAS (CMS), chargino pair production assuming a Bino-like LSP was excluded for \chinoonepm masses up to 420 (200) GeV assuming a W-boson-mediated chargino decay ~\cite{SUSY-2018-32,Sirunyan:2018lul}.
For the scenario assuming a Higgsino-like LSP, \chinoonepm masses below 193 GeV were excluded for mass splittings down to 4.6 GeV, and mass splittings from 1.2 GeV to 30 GeV were excluded at the LEP bounds on \chinoonepm mass by ATLAS ~\cite{SUSY-2018-16}.
Light higgsinos with mass splittings below 1 GeV, preferred by naturalness conditions from the low-energy fine-tuning measures, are still not excluded by the LHC and will be very challenge for the HL-LHC too.

For the Bino-like (Higgsino-like) LSP case, the increased center-of-mass energy of CEPC will extend the search sensitivity for \chinoonepm (\chinoonepm and \ninoone) masses by more than 100 GeV compared to the LEP experiment.
With a cleaner collision environment and better reconstruction and identification efficiency for low-energy particles~\cite{Ruan:2018yrh}, CEPC will have excellent sensitivities to the very compressed electroweakino and sleptons (superpartners of the leptons) search scenarios which are very difficult to search for at the LHC and even HL-LHC.

The FCC-ee and ILC are proposed electron-positron colliders~\cite{Behnke:2013lya,Gomez-Ceballos:2013zzn}.
The CEPC and FCC-ee are both circular colliders designed to be built and operated in several stages with center-of-mass energies from 90 GeV to 350 GeV. The ILC is a linear collider with designed center-of-mass energies from 250 GeV to 1 TeV. 
These colliders will primarily operate with center-of-mass energies at the ZH production threshold, so an energy of 240 GeV is assumed for these studies.
A conservative systematic uncertainty of 5\% is assumed in this paper and is consistent with the LEP results~\cite{ALEPH_syst,DELPH_syst,OPAL_syst}.
The results presented here are expected to be largely independent of the specific detector, trigger and data acquisition choices made, and can be easily considered as a reference for studies of the other two facilities with proper scaling of the luminosity.

\begin{figure}[!hb]
\centering
\includegraphics[width=.45\textwidth]{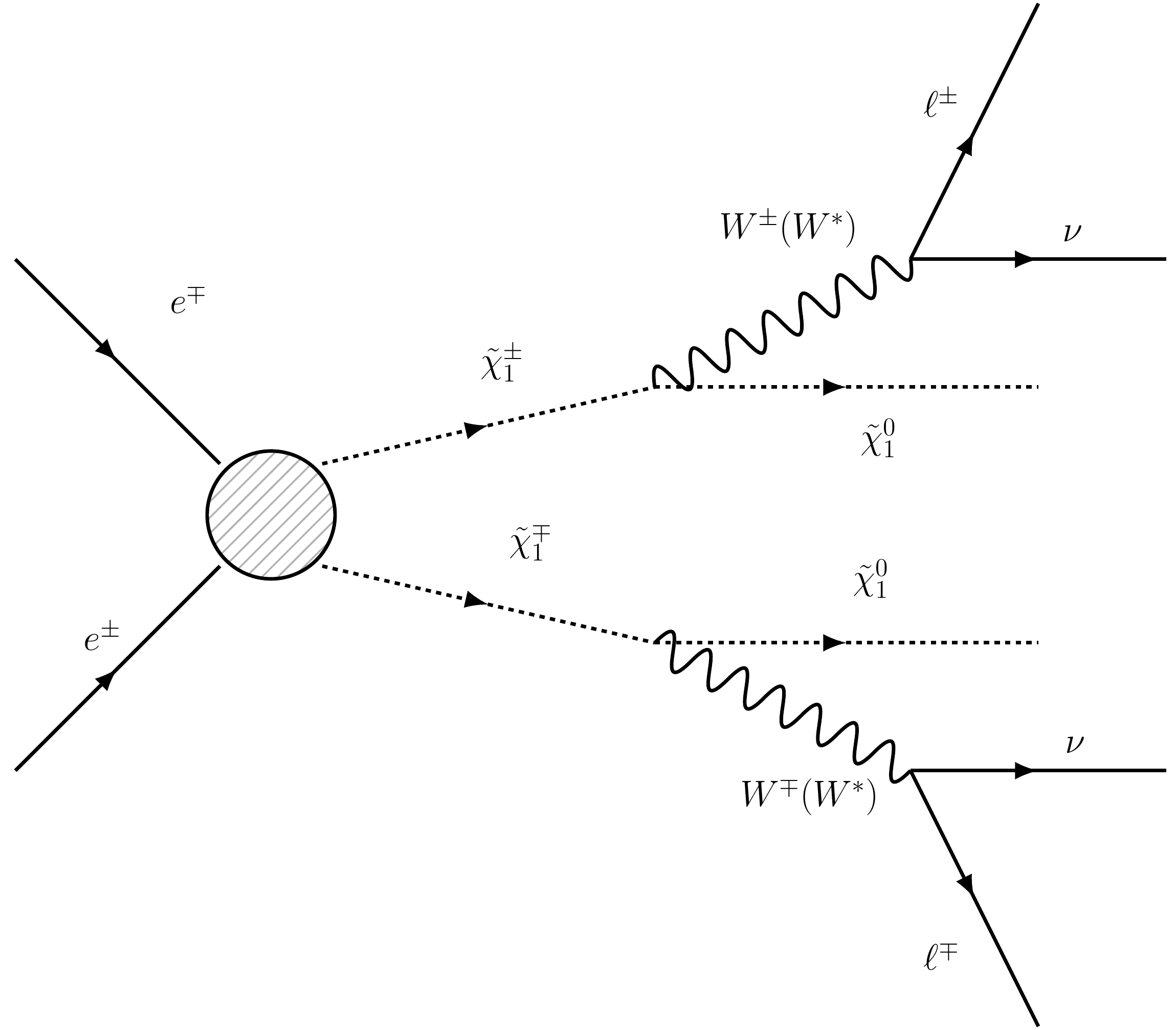}
\caption{Representative diagram illustrating the pair production of charginos and subsequent decay into a two-lepton final state via W bosons.}
  \label{fig:feynmanc1}
\end{figure}

This paper presents studies of the sensitivity to chargino pair production, followed by the chargino decay to a W boson and a Bino-like or Higgsino-like neutralino LSP, as illustrated in Figure \ref{fig:feynmanc1}. 
In order to simplify the analysis, only leptonic W decays are considered. Events with two opposite-sign (OS) muons and significant recoil mass ($M_{recoil}$, the invariant mass of the recoil system against the two muons) are selected for all scenarios.

 \section{Detector, Software and Samples}
 \label{sec:samples}
 The CEPC Conceptual Design Report (CDR) supplies detailed information on the detector and software~\cite{CEPCStudyGroup:2018ghi}.
Two kinds of CEPC detector concepts are proposed.
The baseline concept is a particle-flow oriented detector equipped with an ultra-high granularity calorimeter, a low-material tracker and a 3 Tesla solenoid.
The alternative concept uses a dual readout calorimeter and a 2 Tesla solenoid.
In this analysis, the CEPC baseline concept is used as the detector model.

The following software was used for the study:
Whizard 1.95~\cite{Kilian:2011sepTEPJC} is used to generate the official SM MC simulated event samples.
MadGraph 2.7.3~\cite{Alwall:2014julJHEP} and Pythia 8.244~\cite{Sjostrand:2014zea} are used for the generation of SUSY MC simulated event samples.
MokkaC 0.1.0~\cite{MorasMokka} is used to simulate the interactions between particles and the detector.
Clupatra 00-10~\cite{Gaede_2014} is used to reconstruct tracks from hits in the detector.
The particle flow algorithm Arbor 3.4.2~\cite{Ruan:2018mayTEPJC} is used to reconstruct physics object.
LICH v4~\cite{Yu:2017mpx} based on Multivariate Data Analysis (TMVA)~\cite{hoecker2007tmva} is used for lepton identification.

The \chinoonepm are pair produced from electron-positron collisions and each \chinoonepm decays into a W boson and a \ninoone with a 100\% branching ratio.
No other sparticle is considered in the production or decay.
The signal samples of chargino pair production in the Bino-like LSP case are parametrized as function of the mass of the LSP and \chinoonepm : the lower bound on the chargino mass is set by the LEP limit, while the LSP mass is bound by the mass difference between the chargino and the W boson. 
The \chinoonepm mass is therefore varied in the range 90 - 119 GeV.
The signal samples of chargino pair production in Higgsino-like LSP case are parametrized as function of two SUSY parameters $\mu$ and tan$\beta$.
The $\mu$ varied in the range 90 - 118 GeV, while tan$\beta$ is varied in the range 10 - 60.
Samples containing $5\times 10^5$ ($1\times 10^5$) events are simulated for each signal point with Bino-like (Higgsino-like) LSP. 

Reference points with a \chinoonepm masses of 110 GeV and with \ninoone masses of 1 GeV, 10 GeV or 25 GeV are used for the Bino-like LSP scenario in this paper to illustrate typical features of the SUSY models to which this analysis is sensitive. 
The corresponding leading order (LO) cross section as computed by MadGraph is 2789 fb. 
For the Higgsino-like LSP scenario, three reference points with tan$\beta$ of 30 and $\mu$ of 90 GeV, 106 GeV or 118 GeV are used in this paper, with theoretical cross sections at LO as computed by MadGraph of 1966 fb, 1522.9 fb and 681.2 fb respectively.

This analysis only considers the dominant SM background processes with final states with two leptons (electrons, muons or taus) and significant recoil mass.
The backgrounds are categorized into three types: Higgs processes, two fermions backgrounds and four fermions backgrounds. 
The Higgs processes considered are $\nu\nu H, H\to\tau\tau$.
The two fermions backgrounds considered are $\mu\mu$ and $\tau\tau$ processes. 
The four fermions backgrounds include ZZ, WW, single Z, single W and Z or W mixing processes. 
The samples are normalised to a luminosity of 5.05 ab$^{-1}$.
The cross sections of the dominant background processes are shown in Table \ref{tab:xSecb}.
\begin{table}[!htp]
    \begin{center}
    \begin{tabular}{c|c}
    \hline  \hline
    processes   & cross section [fb] \\
    \hline
    $\tau\tau$ & 4374.94\\
    $\nu\nu H, H\to\tau\tau$ & 3.07\\
    $ZZ$ or $WW \to \tau\tau\nu\nu$ & 205.84\\
    $ZZ \to \tau\tau\nu\nu$ & 9.2\\
    $\nu Z , Z \to \tau\tau$ & 14.57\\
    $ZZ$ or $WW \to \mu\mu\nu\nu$ & 214.81\\
    $ZZ \to \mu\mu\nu\nu$ & 18.17\\
    $WW \to \ell\ell$ & 392.96\\
    $\nu Z , Z \to \mu\mu$ & 43.33\\
    $\mu\mu$ & 4967.58\\
    $e\nu W , W \to \mu\nu$ & 429.2\\
    $e\nu W , W \to \tau\nu$ & 429.42\\
    $ee Z , Z \to \nu\nu$ & 29.62\\
    $ee Z , Z \to \nu\nu$ or $e\nu W , W \to e\nu$ & 249.34\\
    \hline \hline
    \end{tabular}
    \caption{\label{tab:xSecb} Cross-sections of the dominant background processes in CEPC.}
    \end{center}
    \end{table}

 \section{Search for chargino pair production at CEPC}
 \label{sec:searchew}
 The event reconstruction consists of track reconstruction, particle flow reconstruction and compound physics objects reconstruction.
Tracks are reconstructed from the hits in the detector by Clupatra~\cite{Gaede_2014}.
Particle flow reconstruction uses the tracks and the calorimeter hits to reconstruct single particle physics objects.
The output of particle flow reconstruction can be used to reconstruct compound physics objects such as converted photons, taus and jets. 
The identification efficiency of muons is 99.9\% for energies above 2 GeV~\cite{Yu:2017mpx}.
For muons with energies below 1.3 GeV and at the edge of the barrel region or the overlap region of barrel and endcap, the identification efficiency is lower than 90\%.
The recoil system consists of all the particles except the two OS charged muons.
Without considering the beam energy spread, the resolution of the reconstructed recoil mass is between 300 and 400 MeV~\cite{CEPCStudyGroup:2018ghi}.

\begin{figure}[!htp]
  \centering
    \subfigure [Bino LSP case (2 OS $\mu$, both energy $>$ 10 GeV)] {\includegraphics[width=.48\textwidth]{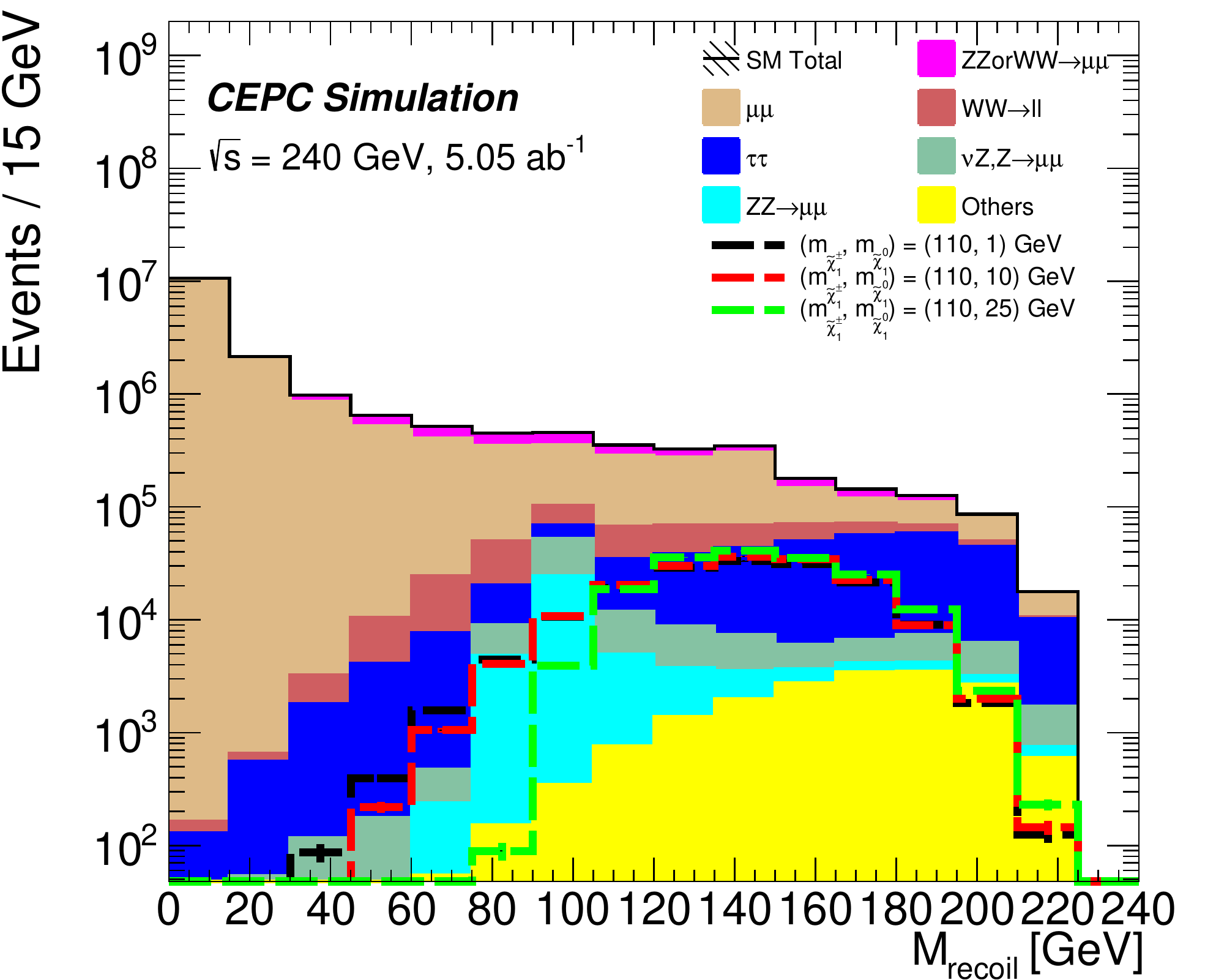}}
    \subfigure [Bino LSP case (2 OS $\mu$, both energy $>$ 10 GeV)] {\includegraphics[width=.48\textwidth]{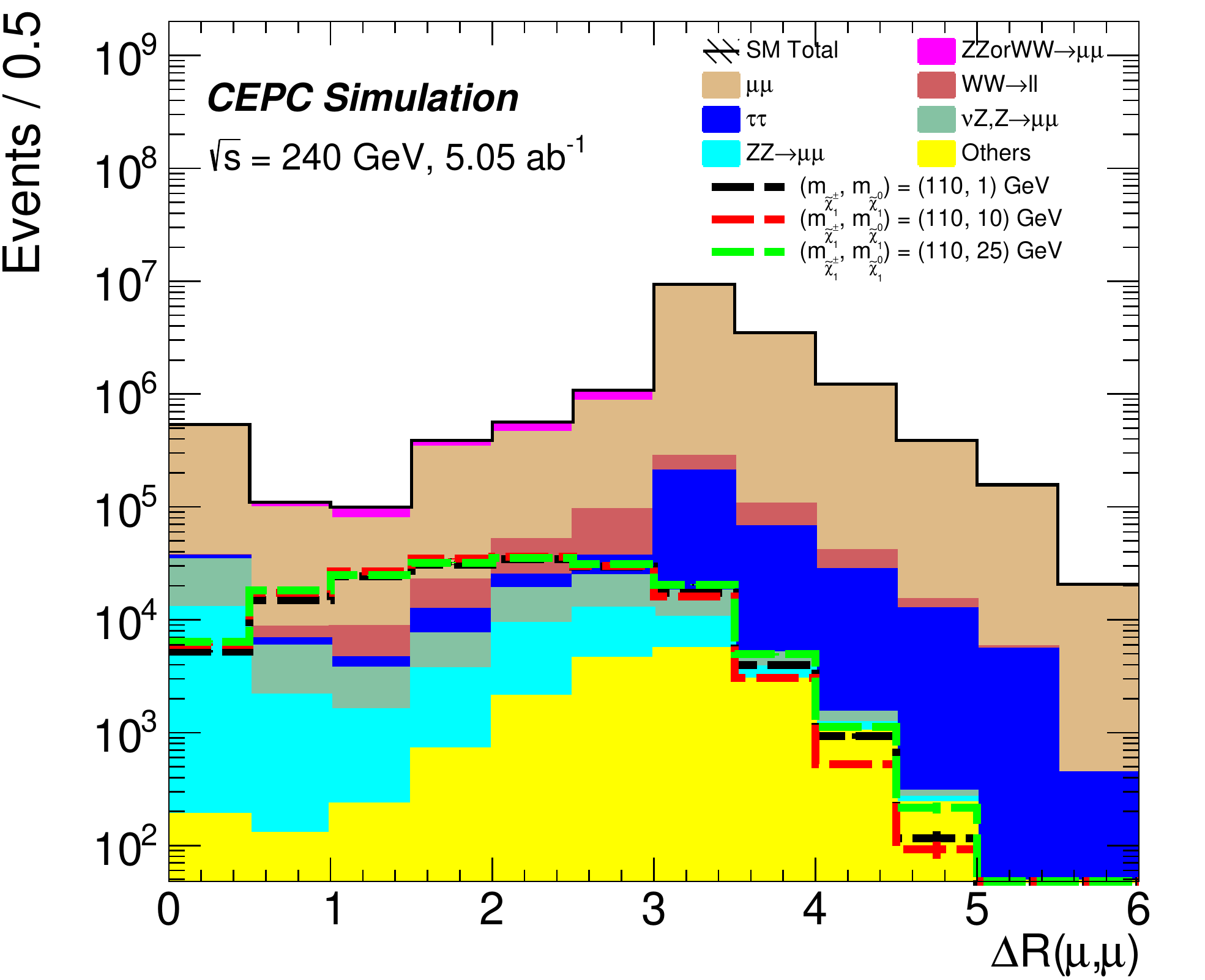}}
    \subfigure [Higgsino LSP case (2 OS $\mu$, both energy $>$ 1 GeV)] {\includegraphics[width=.48\textwidth]{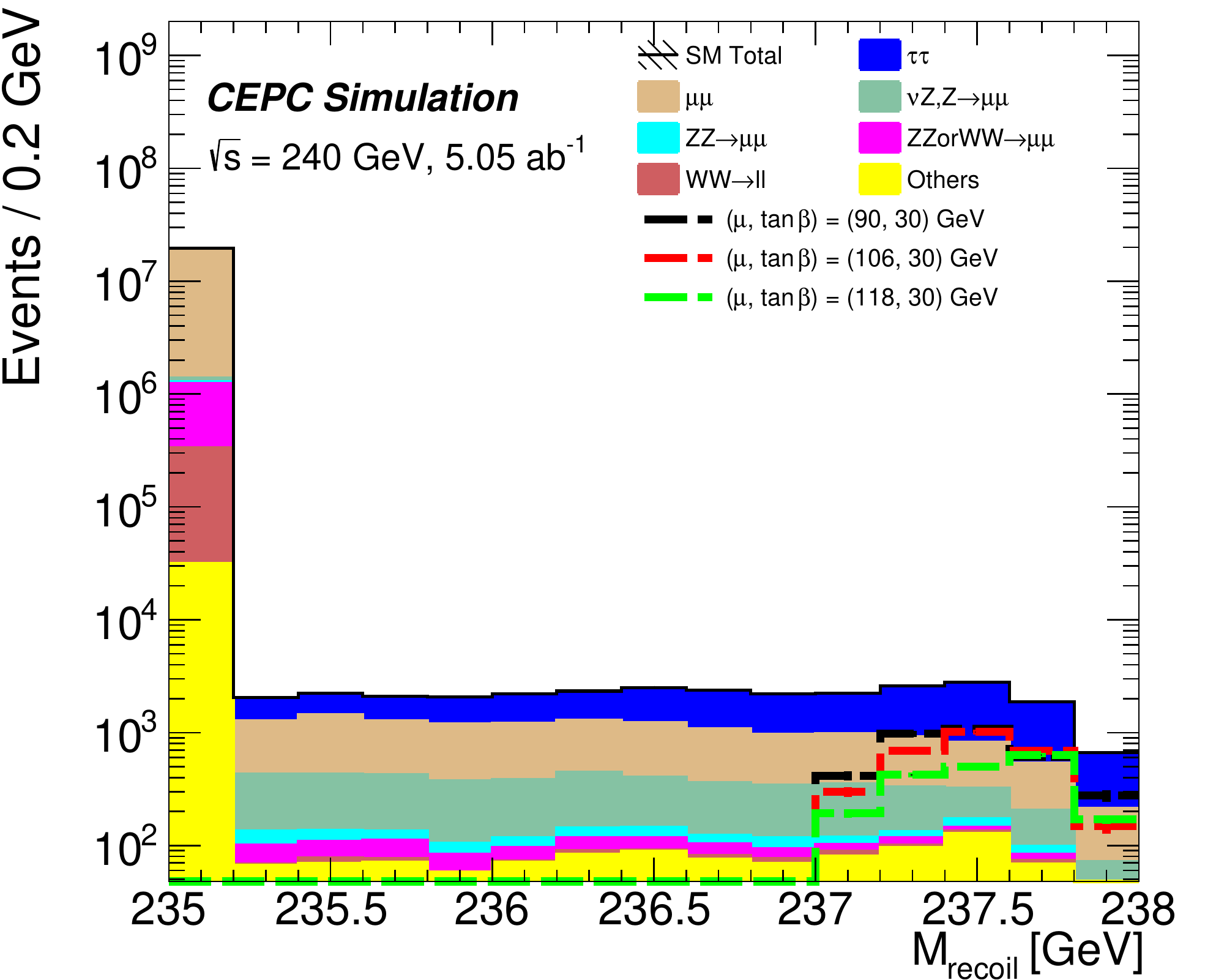}}
    \subfigure [Higgsino LSP case (2 OS $\mu$, both energy $>$ 1 GeV)] {\includegraphics[width=.48\textwidth]{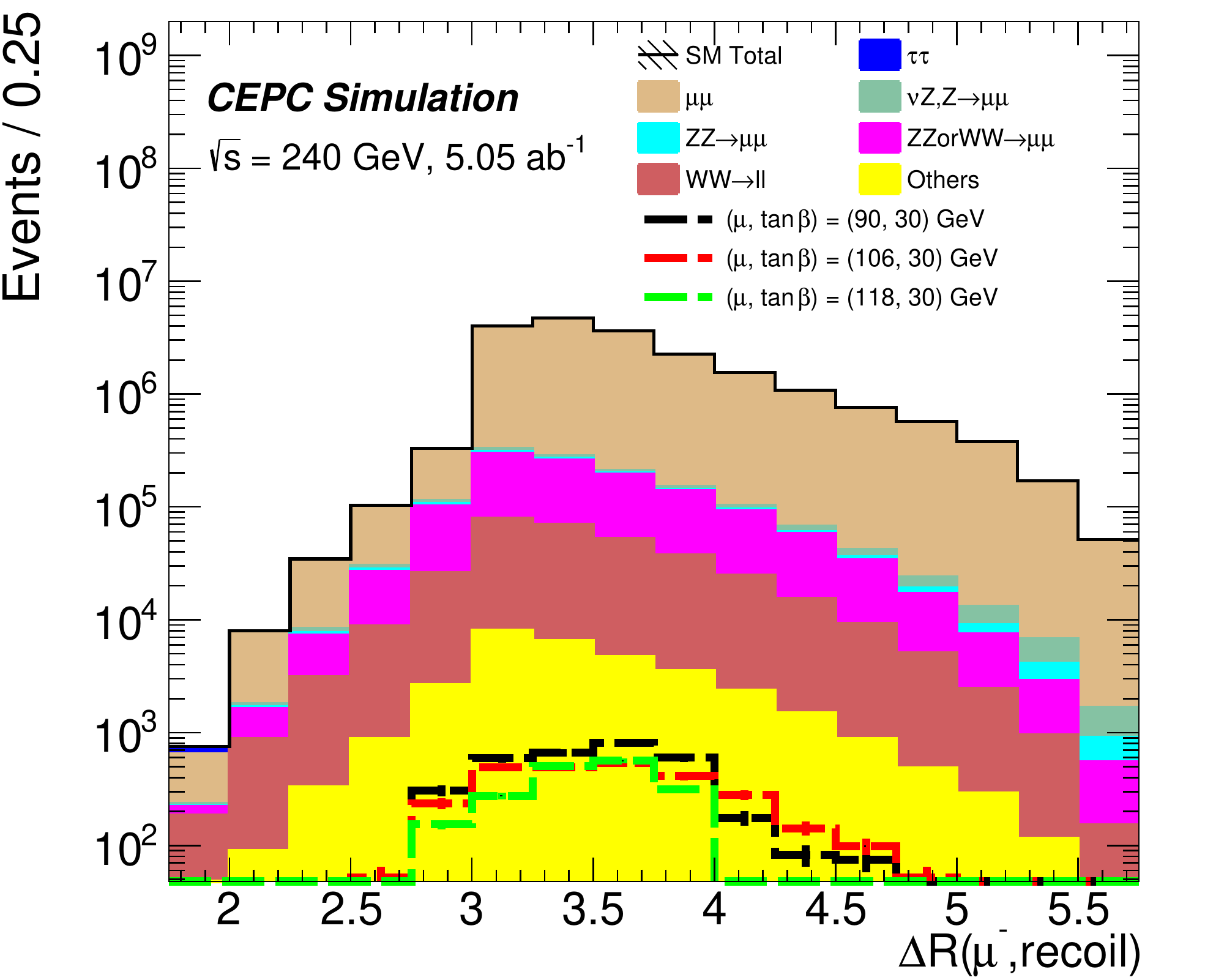}}
    \caption{The kinematic distributions for (top) Bino-like and (bottom) Higgsino-like LSP cases.
    The stacked histograms show the expected SM backgrounds. The hatched bands represent the sum in quadrature of statistical uncertainties of the total SM background. 
    For illustration, the distributions of three SUSY reference points (defined in Sec. \ref{sec:samples}) are also shown as dashed lines. The first bin clearly includes the underflow events and the last bin includes the overflow events.}
    \label{fig:invmass} 
  \end{figure}

The following variables are efficient to discriminate the signal events from SM backgrounds:
\begin{itemize}
\item $\Delta R(\mu^{+},\mu^{-})$, the angular distance between two muons.
\item $\Delta R(\mu^{\pm},recoil)$, the angular distance between the muons and the recoil system.
\item $|\Delta \phi(\mu^{\pm},recoil)|$, the azimuthal distance between the muons and the recoil system.
\item $|\Delta \phi(\mu^{+},\mu^{-})|$, the azimuthal distance between two muons.
\item $E_{\mu^{\pm}}$, the energy of the muons.
\item $P_T^{\mu^{\pm}}$, the transverse momentum of the muons.
\item $M_{recoil}$, the invariant mass of the two neutrinos and two LSPs.
\end{itemize}
The signal regions were optimized and defined based on the above kinematic variables.
Zn~\cite{Cowan:2012ds} was used as a sensitivity reference in the signal region optimization, the defnition of which is shown in formula (\ref{equ:Zn}).
The statistical uncertainty and a 5\% global systematic uncertainty were taken into account in the Zn calculation.
\begin{equation}
    Zn=\left[2\left((s+b)\ln\left[\frac{(s+b)(b+\sigma_b^2)}{b^2+(s+b)\sigma_b^2}\right]-\frac{b^2}{\sigma_b^2}\ln\left[1+\frac{\sigma_b^2s}{b(b+\sigma_b^2)}\right]\right)\right]^{1/2} \label{equ:Zn}
  \end{equation}

Several kinematic distributions after the two OS muon selection with $E_{\mu}$ larger than 10 GeV (1 GeV) are shown in Figure \ref{fig:invmass} (a)(b) (Figure \ref{fig:invmass} (c)(d)), and show the good discrimination power between signal and SM processes.

\subsection{Search for chargino pair production with Bino LSP}
\label{subsec:searchcb}
For the Bino-like LSP scenario, events containing exactly two OS charged muons with energy larger than 10 GeV are selected.
A selection on $\Delta R(\mu^{+},\mu^{-})$ is used to reject background events from $\tau\tau$ process and $\mu\mu$ process.
Events are required to have $P_T^{\mu^{\pm}} > $  30 GeV to suppress WW and Z$\nu$ processes.
Most of the signal events have large recoil mass according to the signal topology as shown in Figure \ref{fig:invmass} (a), so $M_{recoil} > $ 130 GeV is applied to reject $\mu\mu$ and ZZ background events and to increase the signal sensitivity.
The signal region definition is summarized in Table \ref{tab:SRcb}.
\begin{table}[!htp]
  \begin{center}
  \begin{tabular}{c}
  \hline \hline
 Signal Region\\
  \hline
  == 2 muons (OS) \\
  $E_{\mu^{\pm}} > 10 $ GeV\\  
  $0.4 < \Delta R(\mu^{+},\mu^{-}) < 1.6$\\
  $P_T^{\mu^{\pm}} > $  30 GeV\\
  $M_{recoil} > $  130 GeV\\
  \hline \hline
  \end{tabular}
  \caption{\label{tab:SRcb} Summary of signal region selection requirements for the chargino pair production with Bino LSP.}
  \end{center}
  \end{table}

\begin{figure}[!htp]
\centering
  \subfigure [$M_{recoil}$ ($M_{recoil}$ 	\textgreater 130 GeV)] {\includegraphics[width=.48\textwidth]{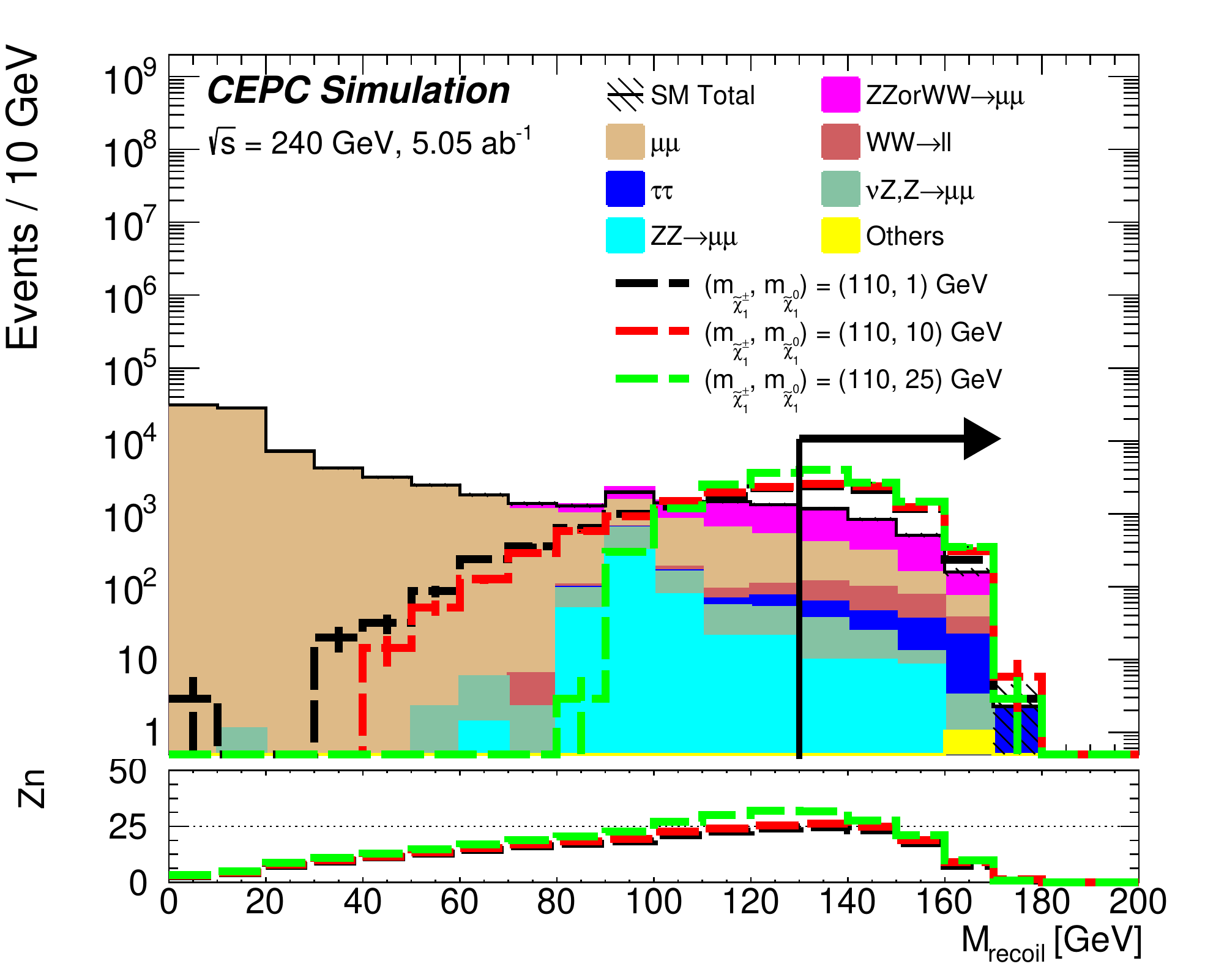}}
  \subfigure [$P_T^{\mu^{-}}$ ($P_T^{\mu^{-}}$ 	\textgreater 30 GeV)] {\includegraphics[width=.48\textwidth]{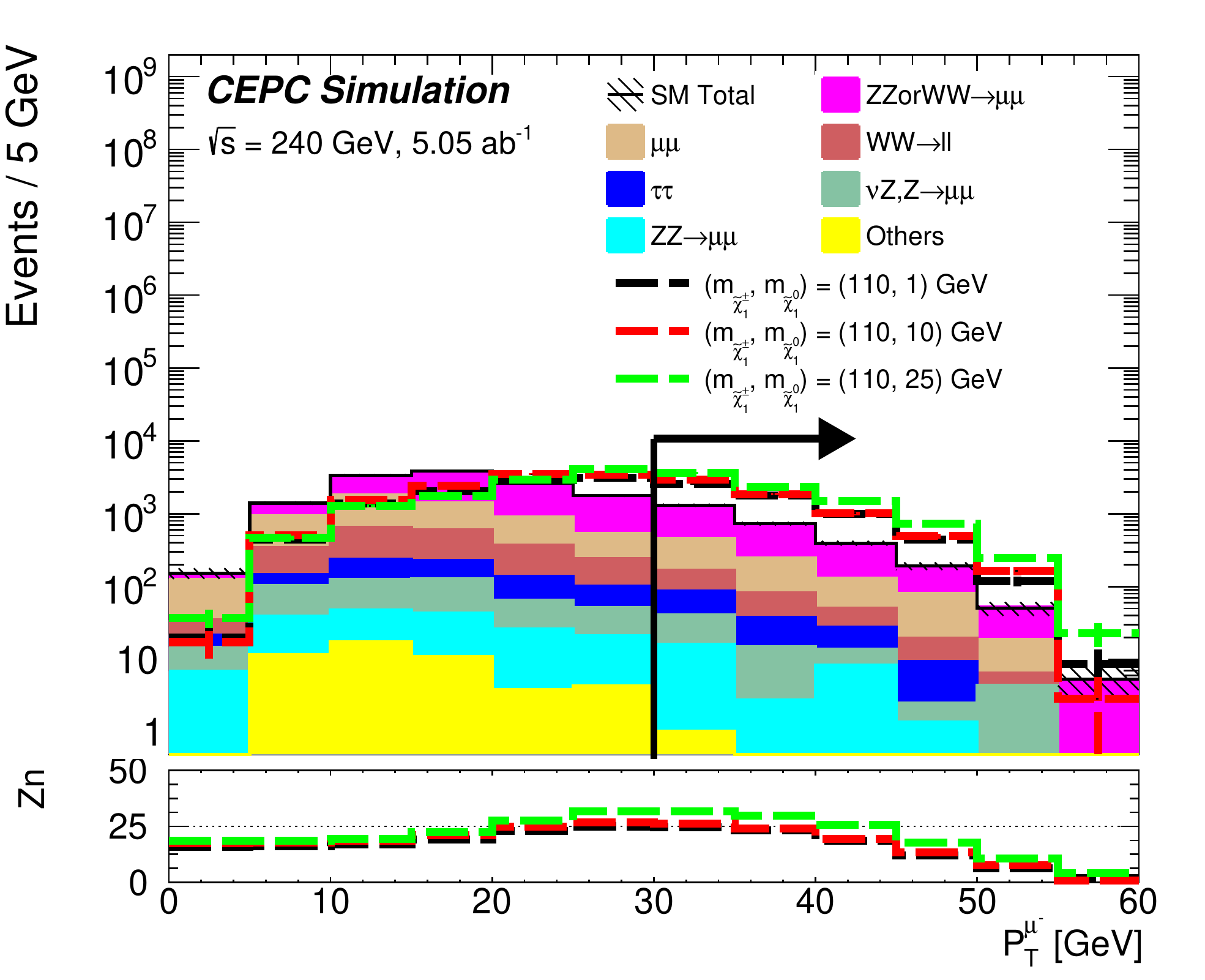}}
  \caption{``N-1'' distributions after signal region requirements for the Bino-like LSP case. All signal region requirements are applied except on the variable shown. The stacked histograms show the expected SM backgrounds. The hatched bands represent the sum in quadrature of statistical uncertainties of the total SM background. For illustration, the distributions from the SUSY reference points (defined in Sec. \ref{sec:samples}) are also shown as dashed lines. The lower pad is the sensitivity Zn calculated with statistical uncertainty and 5\% flat systematic uncertainty. The first bin clearly includes the underflow events and the last bin includes the overflow events.}
  \label{fig:nm1cb}
\end{figure}

\begin{figure}[!htp]

\centering
  \subfigure [systematic uncertainty = 0\%] {\includegraphics[width=.48\textwidth]{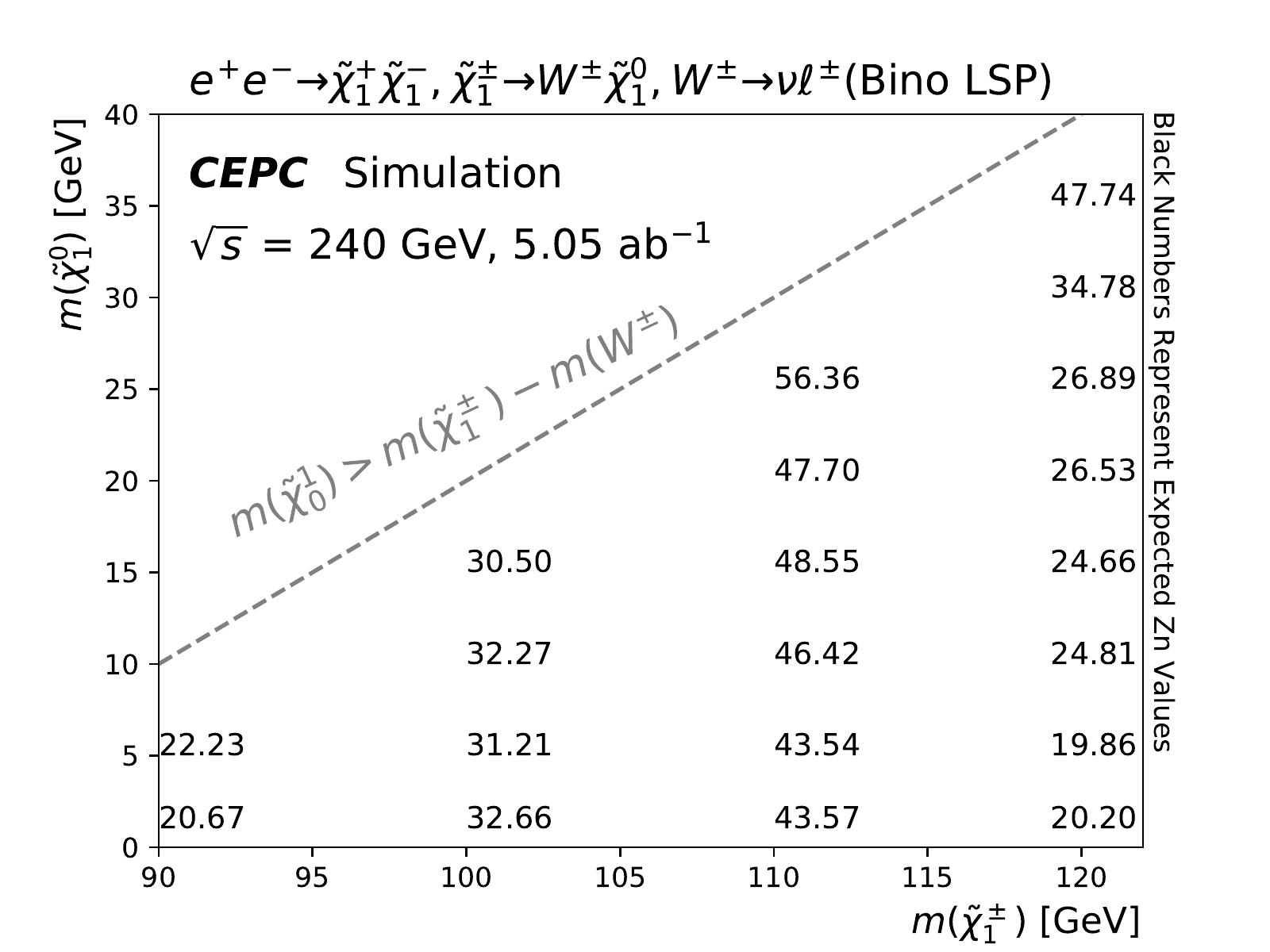}}
  \subfigure [systematic uncertainty = 5\%] {\includegraphics[width=.48\textwidth]{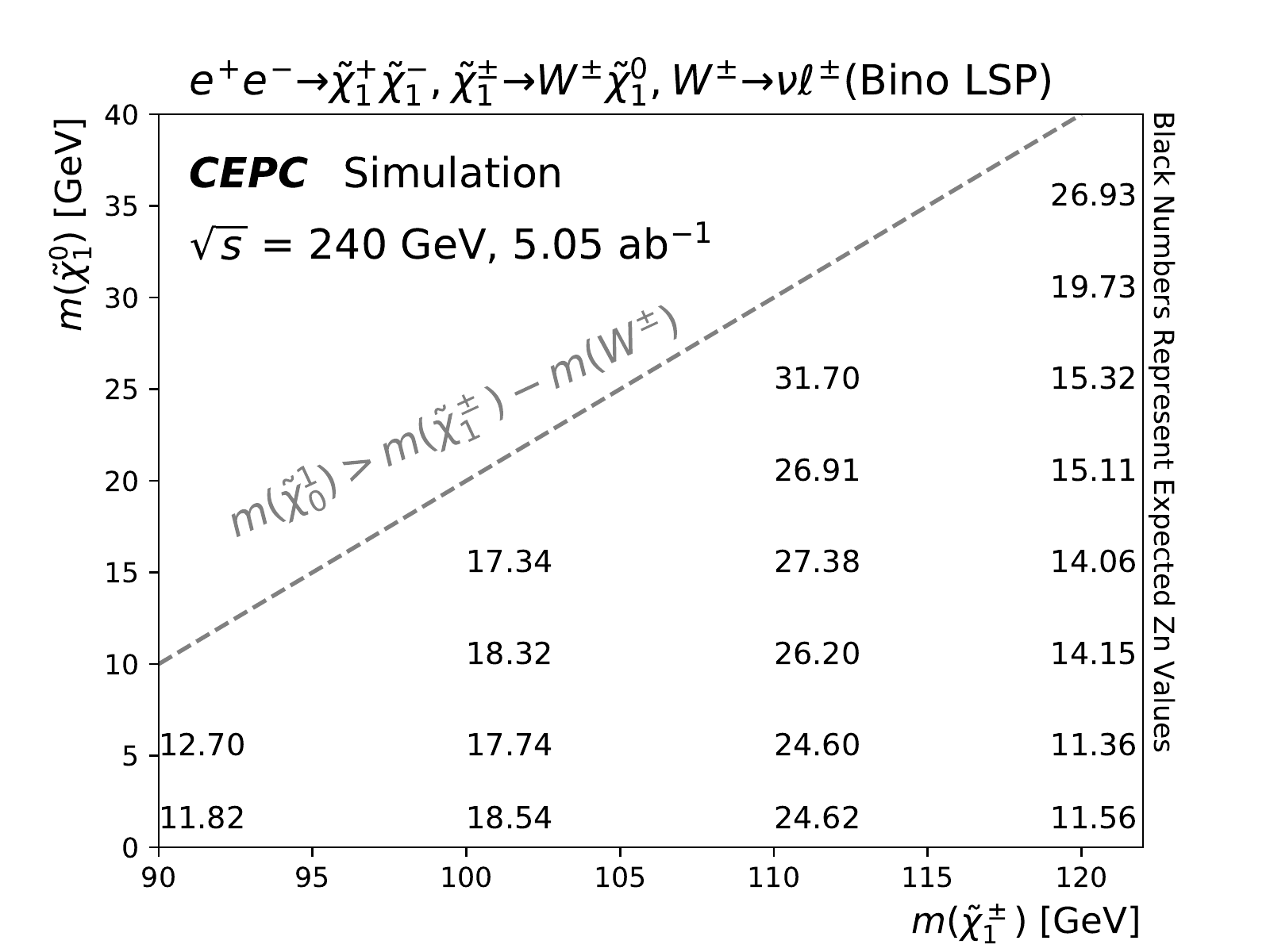}}
  \caption{The expected sensitivities as a function of the \chinoonepm and \ninoone masses for chargino pair production in the Bino-like LSP scenario. Results are made assuming a systematic uncertainty of (left) 0\% or (right) 5\%. Black numbers represent expected Zn values.}
  \label{fig:mapcb}
\end{figure}

The kinematic distributions of $M_{recoil}$ and $P_{T}^{\mu^{-}}$, after applying all signal region requirements except on the shown variable, are given in Figure \ref{fig:nm1cb}.
The "Others" component includes $\nu\nu H$, $H\to\tau\tau$, $ZZ$ or $WW \to \tau\tau\nu\nu$, $ZZ \to \tau\tau\nu\nu$, $\nu Z , Z \to \tau\tau$, $e\nu W , W \to \mu\nu$, $e\nu W , W \to \tau\nu$, $ee Z , Z \to \nu\nu$, $ee Z , Z \to \nu\nu$ or $e\nu W , W \to e\nu$ processes. 
The black line with arrow indicates the signal region selection of this kinematic variable.
The sensitivities shown in the lower pad of the figures are obtained from the cumulative signal and background events calculated in the direction of the cut arrow. 
The event yields from background processes and signal reference points after signal region requirements are shown in Table \ref{tab:numcb}.
The dominant background contributions are from $ZZ$ or $WW \to \mu\mu\nu\nu$, $\mu\mu$ and $WW\to\ell\ell$ processes.
The expected sensitivities as function of the \chinoonepm and \ninoone masses with systematic uncertaintyies of 0\% or 5\% for the Bino-like LSP case are shown in Figure \ref{fig:mapcb}.
The discovery potential can reach up to the kinematic limit of $\sqrt{s}/2$, and is not sensitive to systematic uncertainties. 

\begin{table}[!htp]
\begin{center}
\begin{tabular}{c|c}
\hline \hline
Process & Yields\\
\hline
$ZZorWW\to\mu\mu\nu\nu$ &    1632$\pm$42 \\
$\mu\mu$ &     609$\pm$61 \\
$WW\to\ell\ell$ &     163$\pm$13 \\
$\tau\tau$ &      88$\pm$14 \\
$\nu Z,Z\to\mu\mu$ &   47.9$\pm$7.3 \\
$ZZ\to\mu\mu\nu\nu$ &   27.7$\pm$6.2 \\
Others &  0.74$\pm$0.74 \\
\hline
Total background & 2568 $\pm$ 77 \\
\hline
m ( \chinoonepm , \ninoone ) = ( 110  , 1 ) GeV & 5940 $\pm$ 130 \\
m ( \chinoonepm , \ninoone ) = ( 110 , 10 ) GeV & 6470 $\pm$ 140\\
m ( \chinoonepm , \ninoone ) = ( 110 , 25 ) GeV & 8470 $\pm$ 160 \\
\hline \hline
\end{tabular}
\caption{\label{tab:numcb} The number of events with statistical uncertainties in the signal region for signals and SM backgrounds for chargino pair production with Bino LSP case.}
\end{center}
\end{table}

\subsection{Search for chargino pair production with Higgsino LSP}
\label{subsec:searchch}
For the Higgsino-like LSP scenario, events containing exactly two OS muons with energy above 1.0 GeV are selected.
The cuts on $E_{\mu^{\pm}}$ and $\Delta R(\mu^{\pm},recoil)$ are used to reject background events from $\tau\tau$ processes and Z or W mixing processes.
A requirement of $|\Delta \phi(\mu^{\pm},recoil)| < 2.9$ is used to suppress $\mu\mu$ process, and $|\Delta \phi(\mu^{+},\mu^{-})| <1.4$ is used to suppress the background events with two back-to-back muons, such as from the $\gamma\gamma\to\mu\mu$ process where the photons are produced from the incoming electrons/positrons.
Most of the signal events have large recoil mass according to the signal topology as shown in Figure \ref{fig:invmass} (c), so $M_{recoil} > 237.5$ GeV has been applied to reject $\mu\mu$, Z or W mixing and WW background events and to increase signal sensitivity.
The signal region definition for Higgsino-like LSP case is summarized in Table \ref{tab:SRch}.

\begin{table}[!htp]
  \begin{center}
  \begin{tabular}{c}
  \hline \hline
  Signal Region\\
  \hline
  == 2 muons (OS) \\
  $E_{\mu^{\pm}} > 1.0 $ GeV\\
  3.2  $ < \Delta R(\mu^{\pm},recoil) < $  4.6\\
  $|\Delta \phi(\mu^{\pm},recoil)| < 2.9$ \\
  $|\Delta \phi(\mu^{+},\mu^{-})| < 1.4$\\
  $M_{recoil} > 237.5$ 	 GeV \\
  \hline \hline
  \end{tabular}
  \caption{\label{tab:SRch} Summary of signal region selection requirements for the chargino pair production with Higgsino LSP.}
  \end{center}
  \end{table}
    
\begin{figure}[!htp]
  \centering
    \subfigure [$E_{\mu^{-}}$ ($E_{\mu^{-}} > $ 1.0 GeV)] {\includegraphics[width=.48\textwidth]{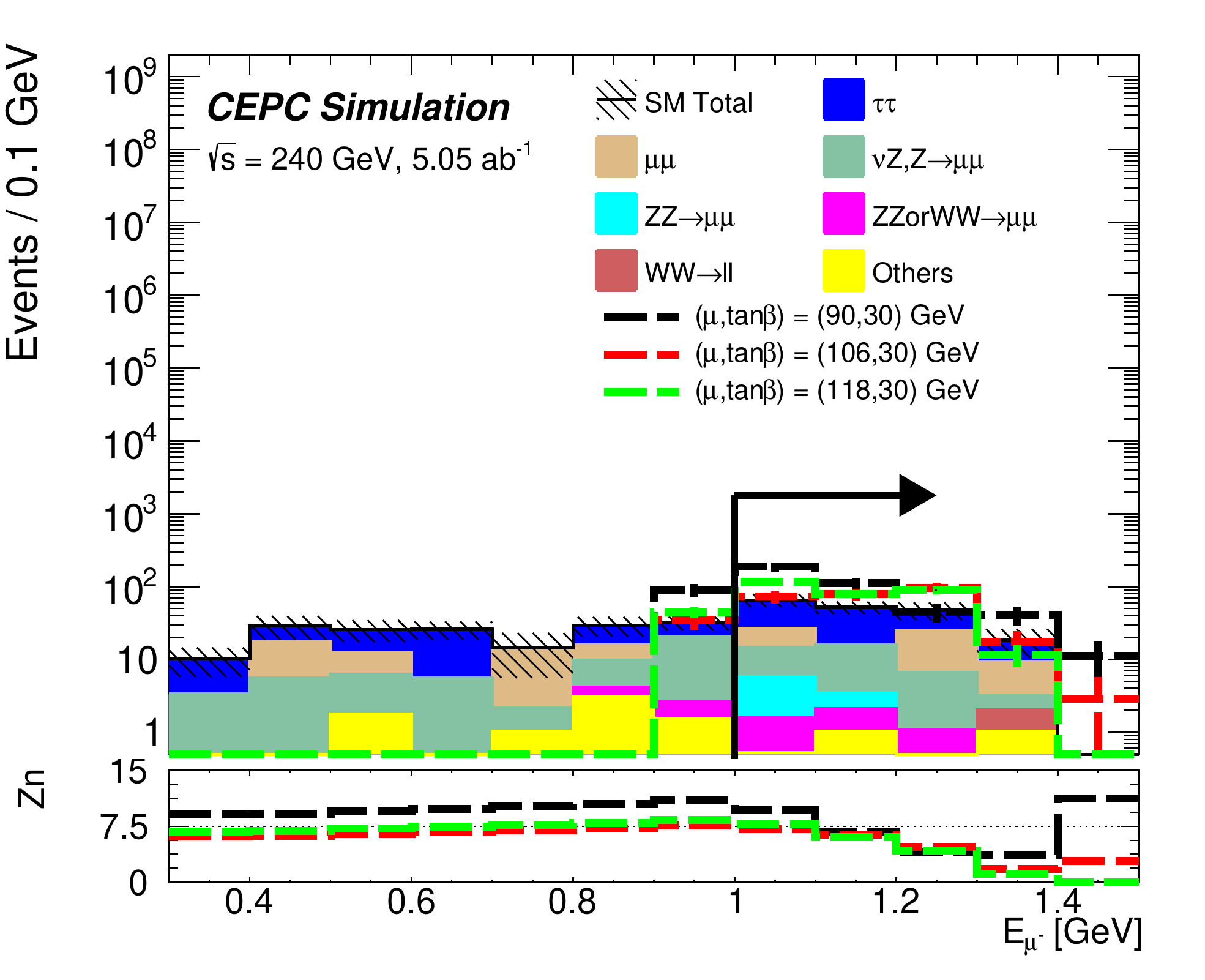}}
    \subfigure [$M_{recoil}$ ($M_{recoil} > $	237.5 GeV)] {\includegraphics[width=.48\textwidth]{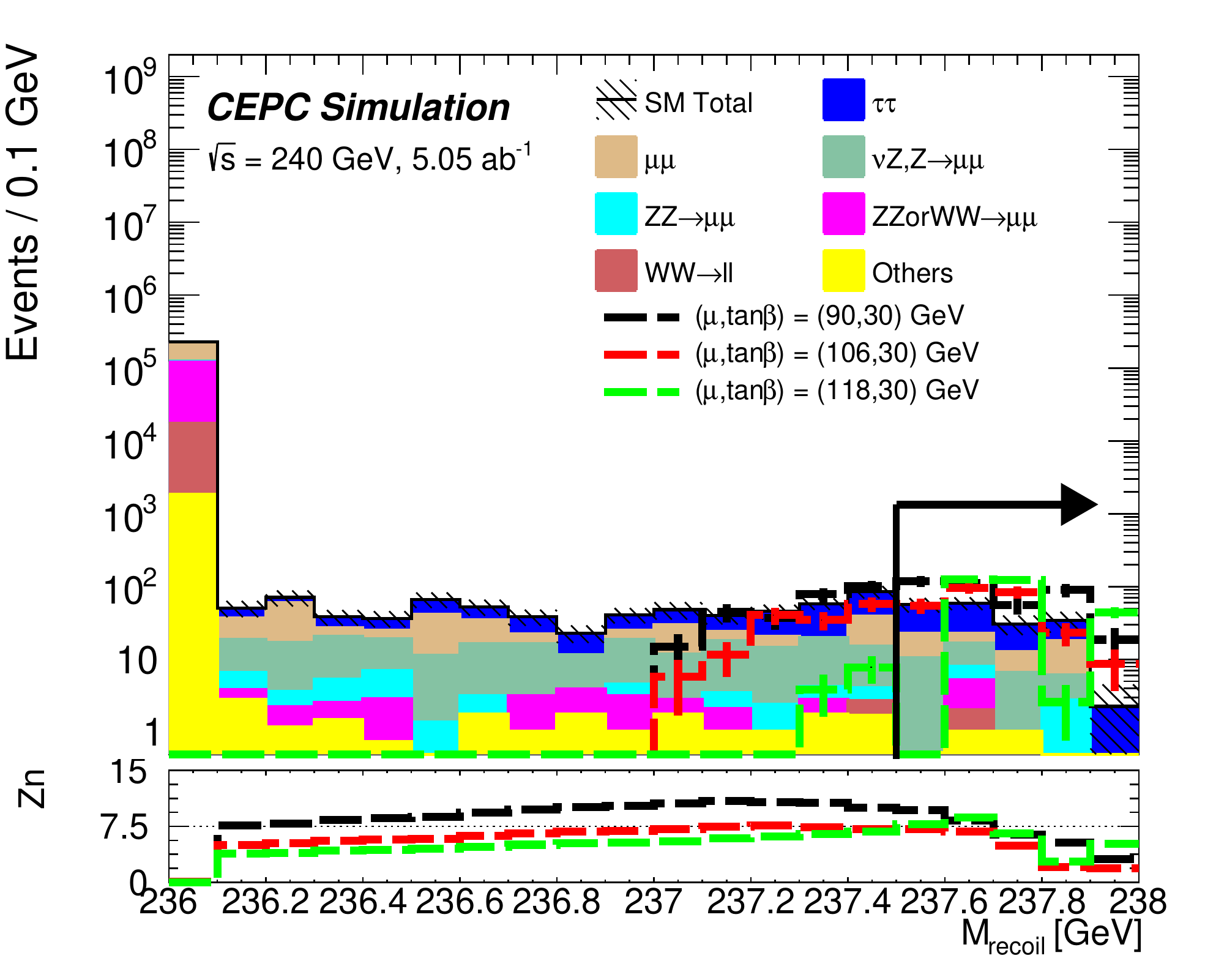}}
    \caption{``N-1'' distributions of used variables after signal region requirements for the Higgsino-like LSP case. All signal region requirements are applied except on the variable shown. The stacked histograms show the expected SM backgrounds. The hatched bands represent the sum in quadrature of statistical uncertainties of the total SM background. For illustration, the distributions from the SUSY reference points (defined in Sec. \ref{sec:samples}) are also shown as dashed lines. The lower pad is the Zn calculated with statistical uncertainty and 5\% flat systematic uncertainty. The first bin includes the underflow events. The first bin clearly includes the underflow events and the last bin includes the overflow events.}
    \label{fig:Nm1ch}
  \end{figure}

\begin{figure}[!htp]
  \centering
    \subfigure [systematic uncertainty =0\%] {\includegraphics[width=.48\textwidth]{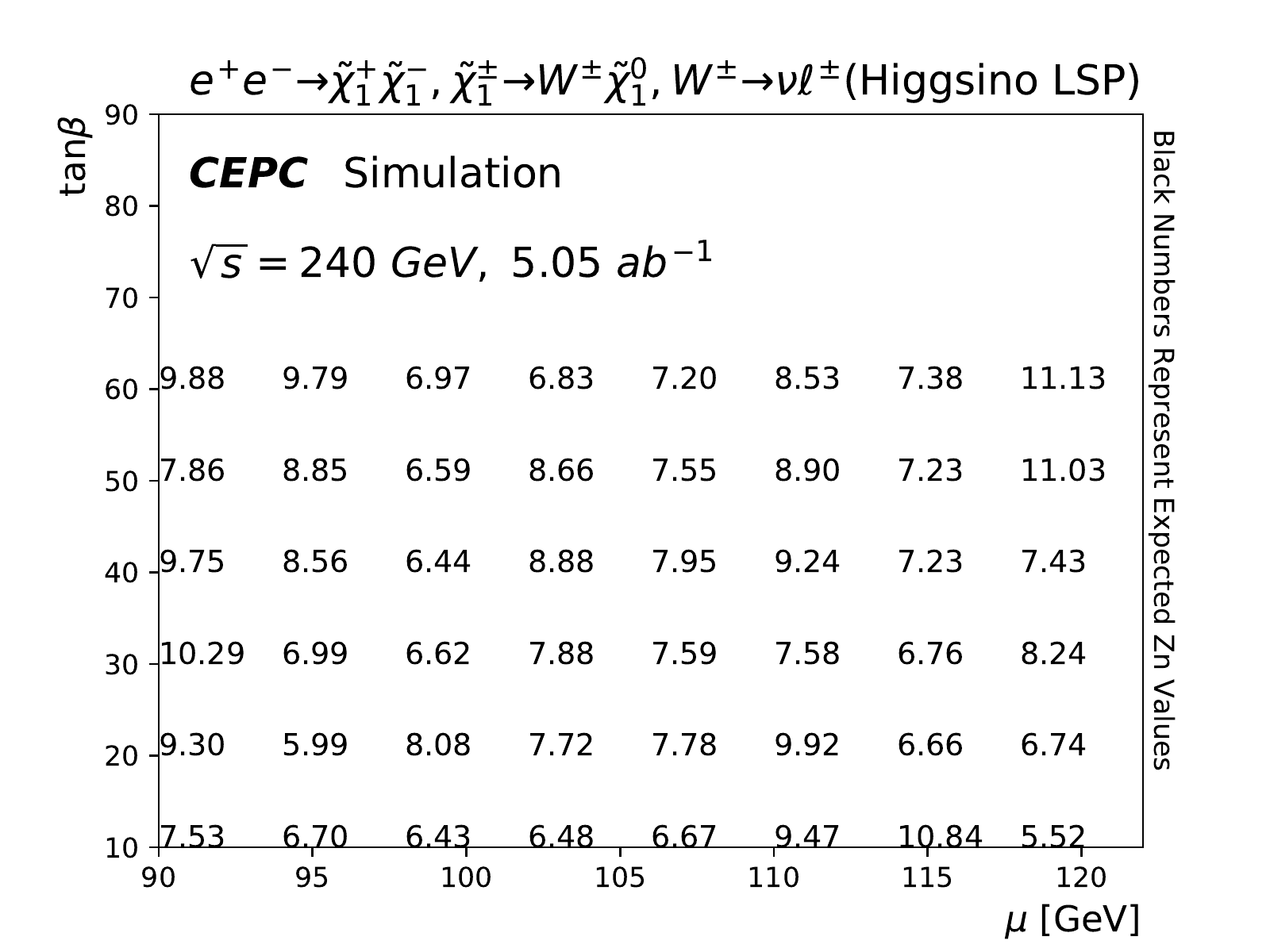}}
    \subfigure [systematic uncertainty =5\%] {\includegraphics[width=.48\textwidth]{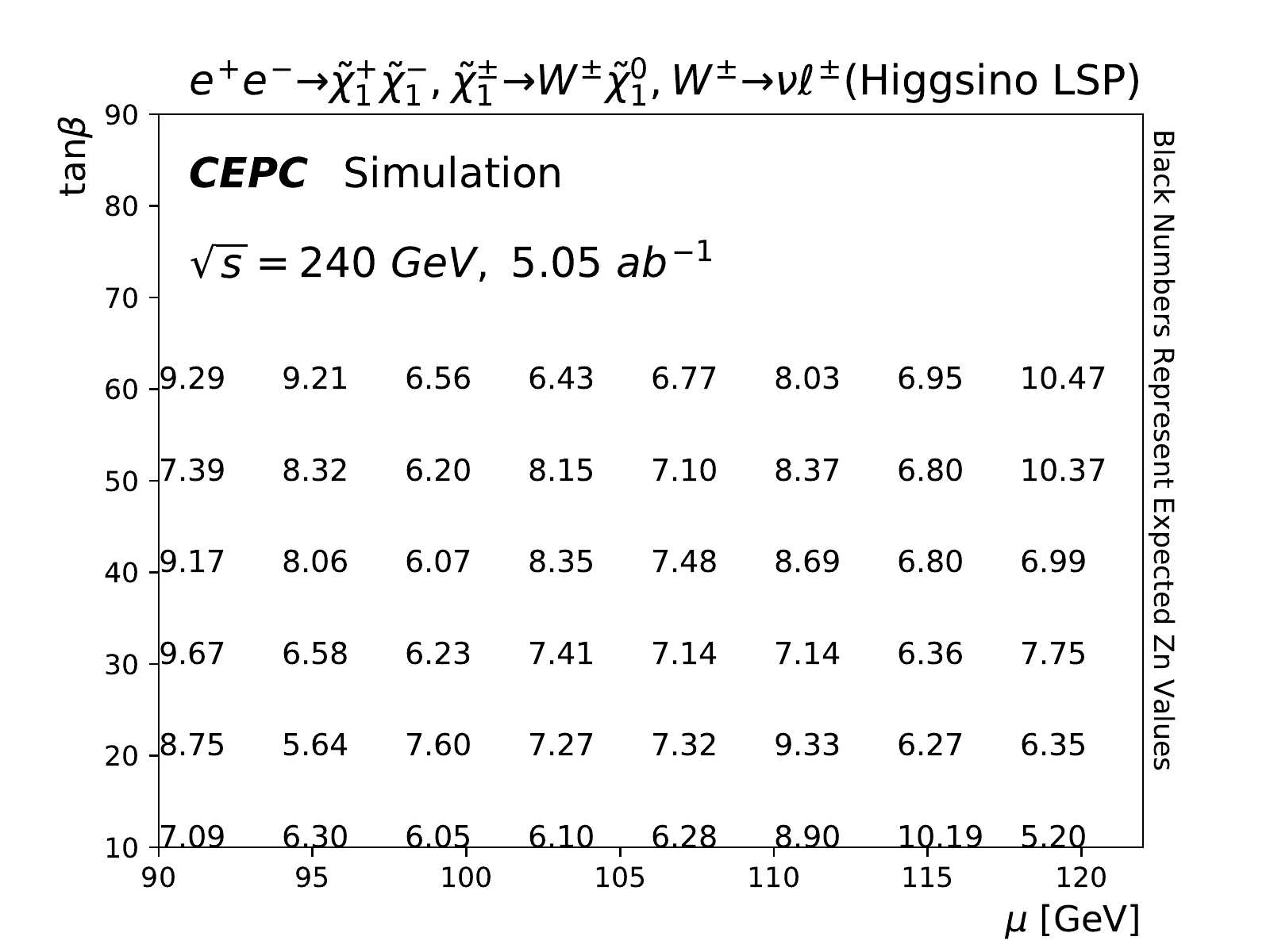}}
    \subfigure [systematic uncertainty =0\%] {\includegraphics[width=.48\textwidth]{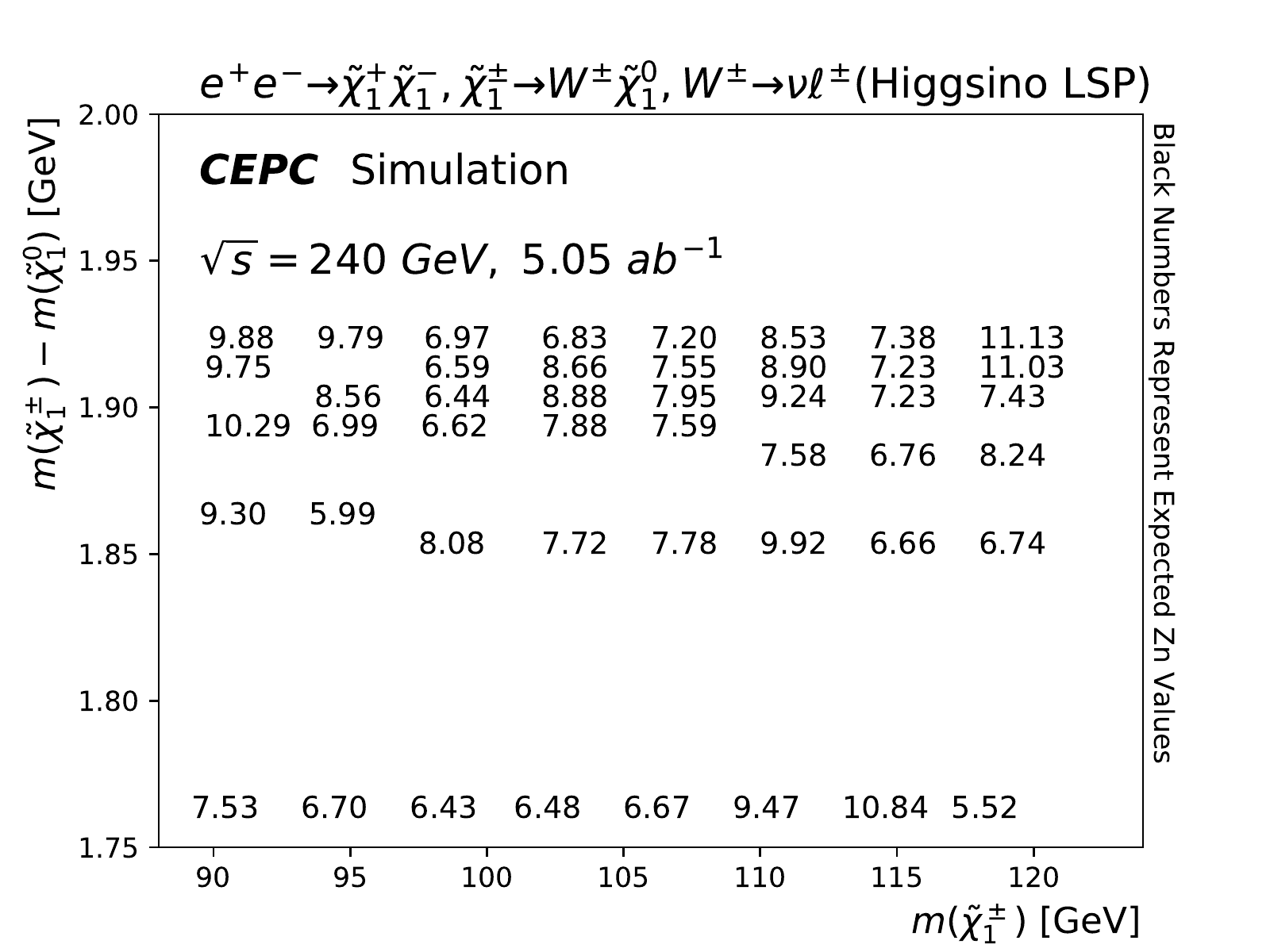}}
    \subfigure [systematic uncertainty =5\%] {\includegraphics[width=.48\textwidth]{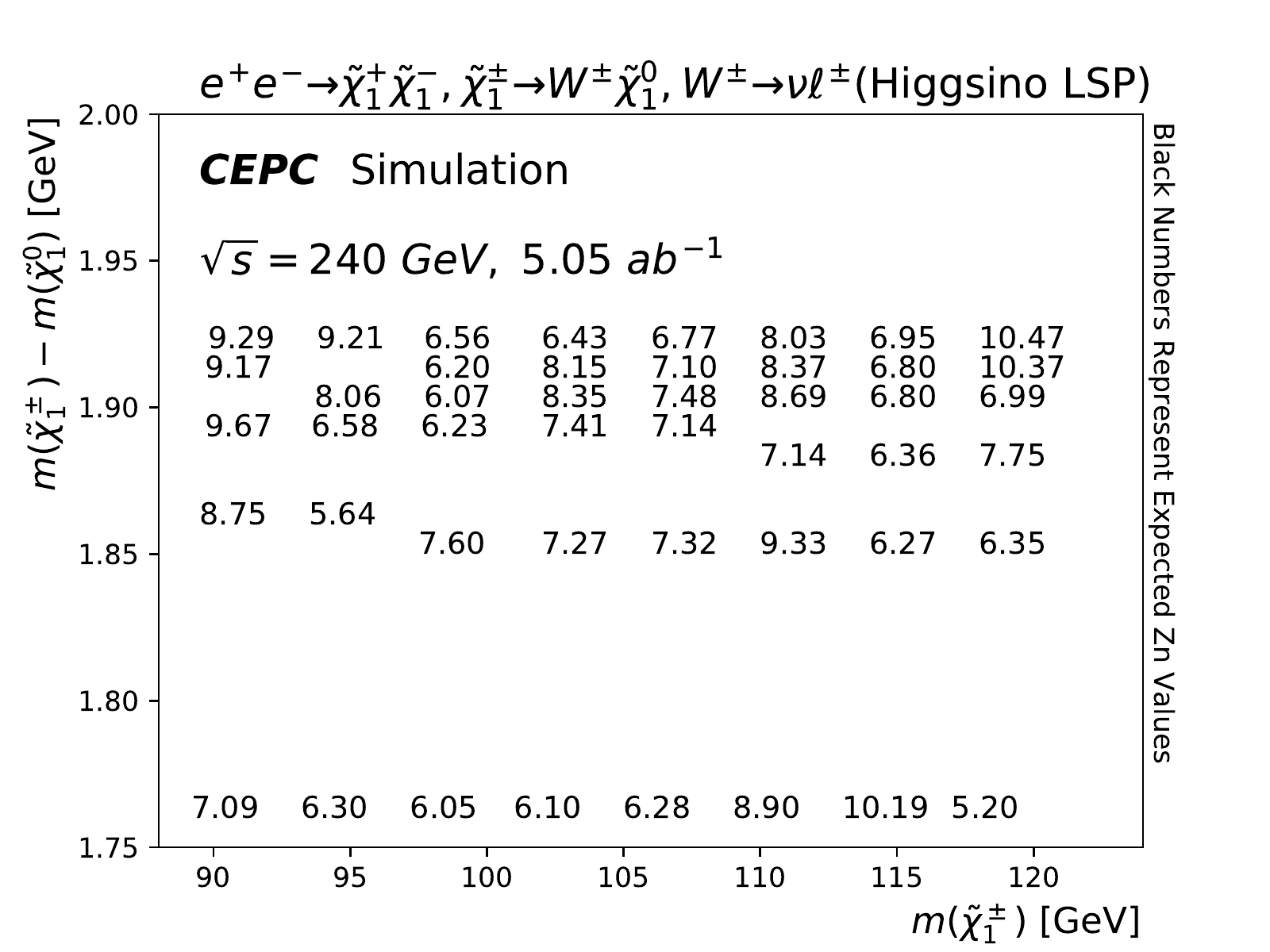}}
    \caption{The expected sensitivities as a function of the \chinoonepm and \ninoone masses for chargino pair production with the Higgsino-like LSP scenario. Results are made assuming a systematic uncertainty of (left) 0\% or (right) 5\%. Black numbers represent expected Zn values.}
    \label{fig:mapch}
  \end{figure}

The kinematic distributions of $E_{\mu^{-}}$ and $M_{recoil}$, after applying all signal region requirements except on the shown variable, are given in Figure \ref{fig:Nm1ch}.
The event yields from background processes and signal reference points after signal region selections are in Table \ref{tab:numch}.
The dominant background contributions are from $\nu Z , Z \to \mu\mu$ and $\tau\tau$ processes.
The expected sensitivities as a function of the \chinoonepm mass and \ninoone mass with systematic uncertainty assumptions of 0\% or 5\% for the Higgsino-like LSP case are shown in Figure \ref{fig:mapch}.
The discovery sensitivity can again reach up to the kinematic limit of $\sqrt{s}/2$, and is not sensitive to systematic uncertainties and mass splitting between NLSP and LSP.
Sensitivity is also expected at lower mass splittings not explored here.

\begin{table}[!htp]
\begin{center}
\begin{tabular}{c|c}
\hline \hline
Processes & Yields \\
\hline
$\tau\tau$ &    107$\pm$16 \\
$\mu\mu$ &     37$\pm$15 \\
$\nu Z,Z\to\mu\mu$ &  27.8$\pm$5.6 \\
$ZZ\to\mu\mu\nu\nu$ &   5.5$\pm$2.8 \\
$ZZ$ or $WW\to\mu\mu\nu\nu$ &   3.2$\pm$1.8\\                                      
$WW\to\ell\ell$ &   1.0$\pm$1.0 \\
Others &   2.6$\pm$1.6 \\
\hline
Total background & 183$\pm$23\\
\hline
( $\mu$ , tan$\beta$ ) = ( 90 GeV , 30 ) &396$\pm$39\\
( $\mu$ , tan$\beta$ ) = ( 106 GeV , 30 )&267$\pm$28\\
( $\mu$ , tan$\beta$ ) = ( 118 GeV , 30 ) &296$\pm$20 \\
\hline \hline
\end{tabular}
\caption{\label{tab:numch} The number of expected events and statistical uncertainties in the signal region for signals and SM backgrounds for chargino pair production in the Higgsino-like LSP case.}
\end{center}
\end{table}

\subsection{Chargino pair production search summary}
\label{subsec:searchsum}
The studies presented in this paper demonstrate that the discovery sensitivity for electroweakinos with both a Bino-like LSP and a Higgsino-like LSP can reach masses up to $\sqrt{s}/2$ at CEPC. The sensitivity can be further improved if considering additional final states (i.e. electron and tau), alternate signal scenarios such as disappearing track or chargino pair production with photon initial-state radiation (ISR), or more complicate analysis techniques (i.e. machine learning etc). 

\begin{figure}[!htp]
  \centering
    \subfigure [] {\includegraphics[width=.48\textwidth]{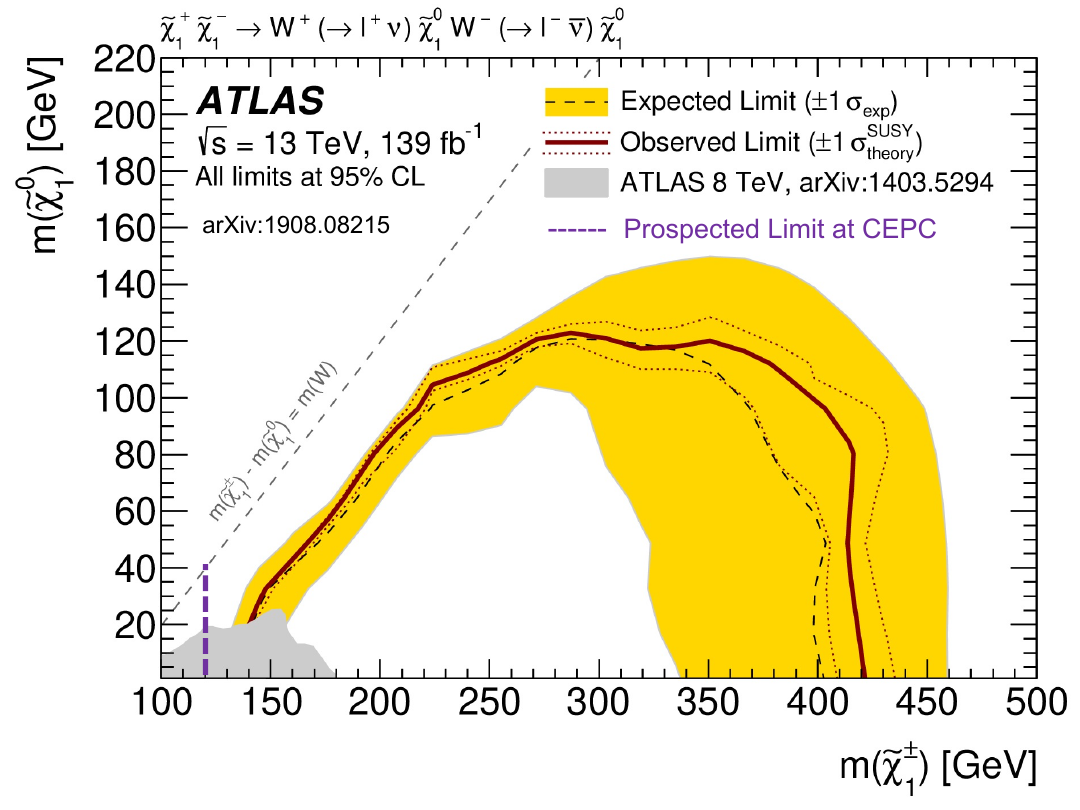}}
    \subfigure [] {\includegraphics[width=.48\textwidth]{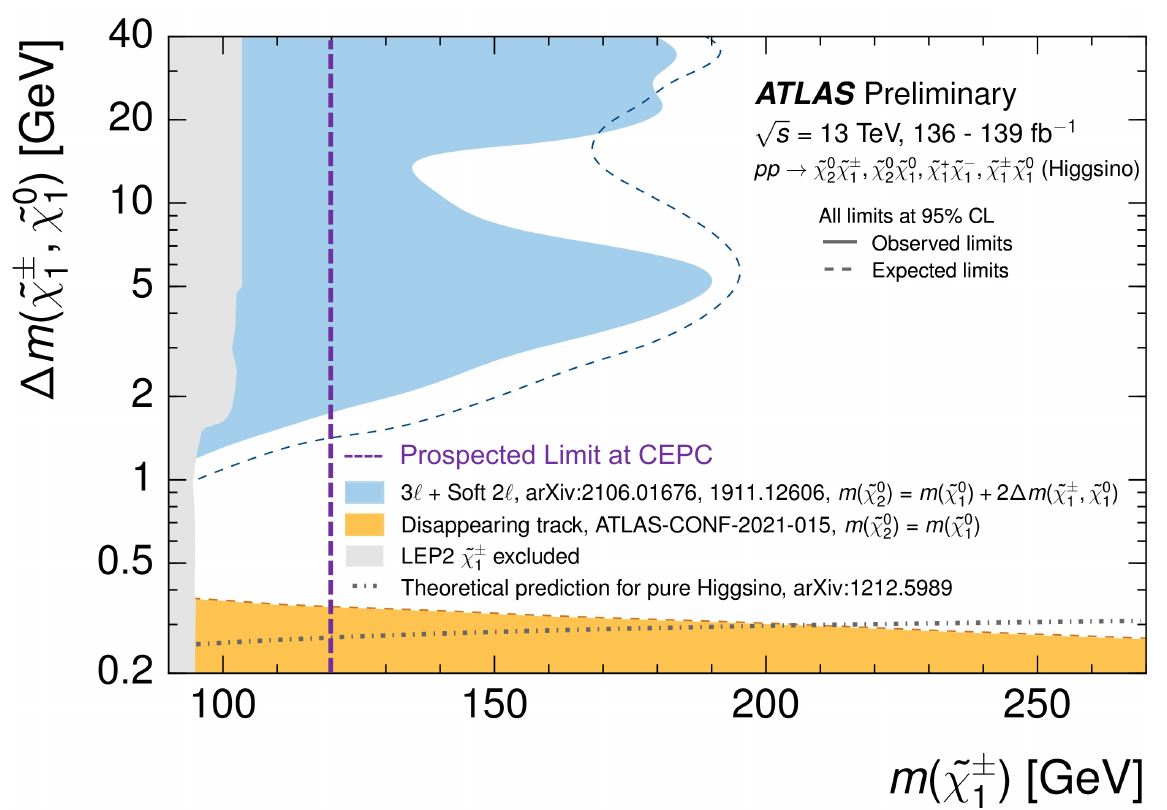}}
    \caption{Observed and expected exclusion limits on SUSY simplified models for chargino-pair production with (a) Bino-like LSP and (b) Higgsino-like LSP obtained by ATLAS. The observed limits obtained by LEP are also shown for Higgsino LSP case in Figure \ref{fig:searchsum} (b), while not shown in Figure \ref{fig:searchsum} (a) since the limit of chargino mass is below 100 GeV. The prospected limits at CEPC are also shown in the dotted purple line for rough comparison. }
    \label{fig:searchsum}
  \end{figure}

The prospective limits at CEPC compared with LEP and LHC results are shown in Figure \ref{fig:searchsum}. 
In general, the sensitivity at CEPC is mainly constrained by the center-of-mass energy and is not strongly dependent on the mass splitting between NLSP and LSP, allowing CEPC to cover very compressed electroweakinos that are very difficult to search for at the LHC due to the reduced acceptance and reconstruction efficiency of the soft leptons.

 \pagebreak
 \clearpage

 \section{Conclusion}
 \label{sec:conclusion}
 Prospective searches for chargino pair production via W boson decay in scenarios with a Bino-like LSP or Higgsino-like LSP are performed at CEPC using simulated samples.
With the assumption of a 5\% systematic uncertainty, the discovery sensitivity in electroweakinos can reach up to $\sqrt{s}/2$.
The choice of systematic uncertainty has no large impact on the discovery sensitivity.
The sensitivity can be further improved by considering more final states, additional signal scenarios or more complex analysis techniques.
CEPC can therefore extend the sensitivity to new Supersymmetry particles, and possible dark matter candiates, of masses up to 120 GeV ~\cite{LEPchargino,Heister:2002mn,Abdallah:2003xe,Chemarin:571780,CERN-OPAL-PN-464,CERN-OPAL-PN-470}.
For the Higgsino-like LSP scenario, CEPC has a very good sensitivity to the compressed regions with a small mass splitting between \chinoonepm and \ninoone, which for the ATLAS and CMS experiments are very difficult to reach~\cite{SUSY-2018-16}.
The result of this research is based on the CEPC expectations but can also be considered as a reference for other lepton collider experiments at a center-of-mass energy close to 240 GeV, such as ILC~\cite{Behnke:2013lya} and FCC-ee~\cite{Gomez-Ceballos:2013zzn}. 

 \section{Acknowledgements}
 \label{sec:acknowledgements}
 The authors thank Cheng-dong Fu, Gang Li, and Xiang-hu Zhao for the simulation code they provided.

\bibliographystyle{spphys} 
\bibliography{EWK}
\end{document}